\documentclass[aps,prb,twocolumn,showpacs,amsmath]{revtex4}
\usepackage{latexsym}
\usepackage{amssymb}
\usepackage{graphicx}
\usepackage{bm}
\usepackage[english]{babel}
\usepackage[latin1]{inputenc}
\usepackage{amsmath}
\usepackage{amsfonts}

\newcommand{\bq}{\begin{equation}}
\newcommand{\eq}{\end{equation}}
\newcommand{\bn}{\begin{eqnarray}}
\newcommand{\en}{\end{eqnarray}}

\begin{document}

\title{Full counting statistics of Kondo-type tunneling in a quantum dot: the fluctuation effect of Slave-Boson field}

\author{Bing Dong}
\affiliation{Department of Physics, Shanghai Jiaotong University,
800 Dongchuan Road, Shanghai 200240, China}

\author{G. H. Ding}
\affiliation{Department of Physics, Shanghai Jiaotong University,
800 Dongchuan Road, Shanghai 200240, China}

\author{X. L. Lei}
\affiliation{Department of Physics, Shanghai Jiaotong University,
800 Dongchuan Road, Shanghai 200240, China}

\begin{abstract}

We study the full counting statistics (FCS) of electron tunneling through a multi-terminal quantum dot in the Kondo regime within 
the slave-boson mean field theory. By employing the A.O. Gogolin and A. Komnik's method of calculating the FCS generating 
function based on the nonequilibrium Green's function [Phys. Rev. B {\bf 73}, 195301 (2006)], we obtain the counting field 
$\lambda$-dependent self-consistent equations for the mean values of the slave-boson fields and the explicit expression for the 
derivative of the adiabatic potential of the system with respect to the counting fields. Performing perturbative expansion to the 
first order of $\lambda$, we find an extra contribution to the shot noise due to the bias-induced Bose field fluctuation, and 
then confirm that the nonequilibrium particle number fluctuation plays an important role in the current noise of the Kondo dot: 
enhancement of the current auto-correlation and a positive current cross-correlation.

\end{abstract}

\date{\today}


\maketitle

\section{Introduction}

Recently, the full counting statistics (FCS)\cite{Nazarov,Blanter} of charge transport in mesoscopic system has become an active 
topic of experimental\cite{Lu,exp1,Fujisawa,exp2,exp3,exp4} and 
theoretical\cite{Levitov,Shelankov,Levitov1,Belzig,Pilgram,Taddei,Lorenzo,Bagrets,Belzig1,Kielich,Urban,Imura,Flindt,Braggio,Emar
y,Dong1} investigation. This concept was first proposed by Levitov and Lesovik\cite{Levitov} to describe the whole probability 
distribution of transmitted charge during a fixed time interval in a noninteracting mesoscopic conductor. It is therefore 
believed that the FCS contain full information about electron correlations that can not be obtained via measuring the average 
current alone.\cite{Nazarov,Blanter} Since then FCS has been studied using the scattering matrix theory in a variety of systems, 
for example, normal and superconductor hybrid tunneling junctions,\cite{Shelankov,Levitov1,Belzig} chaotic quantum dots 
(QDs),\cite{Pilgram} entangled electrons,\cite{Taddei} and spin current.\cite{Lorenzo} Moreover, for taking account of the effect 
of electron-electron interaction, a general scheme has been developed to evaluate the FCS of the Coulomb-blockade QD in the 
framework of a quantum master equation approach.\cite{Bagrets} This scheme has been then applied to analyze the FCS in a 
multilevel QD,\cite{Belzig1} a coupled QDs,\cite{Kielich,Urban} and a single-molecule magnet.\cite{Imura} In combination with 
standard Rayleigh-Schr\"odinger perturbation theory, this approach has been further improved for investigation of the FCS in 
nano-electro-mechanical systems\cite{Flindt} and even for consideration of non-Markovian dynamics.\cite{Braggio} Very recently, 
finite-frequency FCS has been explored in an interacting QD.\cite{Emary} Besides, one of the authors has extended the famous 
MacDonald's formula to calculate the FCS for the nonequilibrated-vibration-assisted tunneling in a molecular QD.\cite{Dong1}

Over last decade, since its experimental discovery in nanoscale devices the Kondo effect and related physics have attracted 
enduring attention.\cite{Goldhaber} This is owing to the impressive advantage of an artificial atom, easy tunability of system 
parameters in a considerably wide range and full controllability of external circumstances, which facilitate a new realm of 
research, the nonequilibrium Kondo physics in transport measurements.\cite{Franceschi,Paaske,Zarchin} For instance, a  
bias-voltage-induced splitting of the Kondo peak in the local density of states of the QD has been experimentally observed in a 
three-terminal transport setup.\cite{Franceschi}  
It is therefore desirable to explore the FCS of electron tunneling passing through a QD connected to multi-terminal in the Kondo 
regime. However, those approaches mentioned above are not applicable for this purpose.

Actually, many attempts have been made in the literature to study the zero-frequency shot noise, i.e. the second current cumulant 
of the FCS, in a Kondo-QD.\cite{Zarchin} Hershfield\cite{Hershfield} calculated the zero-frequency current noise using 
perturbation theory in the Hartree approximation based on the Green function (GF) approach. He found the interaction can either 
enhance or reduce the shot noise. Yamaguchi and Kawamura\cite{Yamaguchi} performed a complementary analysis by choosing the 
tunneling term of the Hamiltonian as the perturbation parameter and revealed a large suppression as comparison with the Poisson 
value. Ding and Ng\cite{Ding} calculated the frequency-dependent shot noise by employing the equation-of-motion method and Ng's 
ansatz for the correlation GF. Their results also demonstrated suppression of the shot noise below the noninteracting value. 
However these results are all only valid at high temperature, $T>T_{\rm K}$ ($T_{\rm K}$ is the Kondo temperature), and weak 
Kondo correlations. For analyzing the shot noise in the case of strong Kondo correlation, Dong and Lei\cite{Dong2} employed the 
finite Coulomb interaction $U$ slave-boson mean-field theory (SBMFT) to calculate the current-current correlation function at 
zero temperature based on the nonequilibrium GF (NGF) technique and Wick's theorem. Later, L\'opez and his coworkers utilized the 
same theoretical framework with the infinite-$U$ version of SBMFT to study the shot noise in a single QD with ferromagnetic 
leads\cite{Lopez1} and multi-terminals,\cite{Sanchez} and a coupled QD system.\cite{Lopez2} Very recently, this method has been 
used to evaluate the shot noise in the parallel coupled QDs to distinguish between the spin and orbital Kondo effect.\cite{Kubo} 
It should be noted that the fluctuations of slave Bose fields are all neglected in the calculation of current correlation 
function.\cite{Dong2} Meir and Golub performed a exhaustive calculation of the shot noise\cite{Meir} based on the noncrossing 
approximation (NCA). However, they just substituted the resulted NCA propagators into the current correlation function derived 
from the {\em noninteracting} electron presumption.

For applying the SBMFT to explore the FCS, one must overcome a technical difficulty: how to evaluate all higher-order terms of 
current correlation functions one-by-one according to Wick's theorem. Even though this is not impossible from theoretical point 
of view (for example, the third cumulant has been derived using Wick's theorem\cite{Golub}), it is actually not practically 
executable. Fortunately, a successful solution to this problem has been given in Ref.\,\onlinecite{Gogolin} done by Gogolin and 
Komnik. In their seminal work,\cite{Gogolin} they introduced a fictitious {\em measuring field} $\lambda$ in the tunneling 
Hamiltonian to count an electron when it goes through the system under studied in the direction of current and assumed it as a 
slowly-varying quantity in time (this is valid since it is set to be equal to zero in the final concrete 
calculations).\cite{Levitov1} This presumption facilitates an adiabatic expansion of the generating function of the cumulants of 
charge current distribution to the first order. By doing so, the calculation of generating function is then transferred to the 
calculation of a so-called adiabatic potential, which can be completely determined by the well defined single-particle Keldysh 
GFs. At the end, they obtained a {\em generic formula} for the generating function of the cumulants expressed only in terms of 
the local Keldysh GFs of the central region (and the tunnel-coupling and Fermi functions of the leads), as done in the 
Meir-Wingreen current formula,\cite{Meir2} except for the presence of the counting field. The advantage of this Hamiltonian 
approach is that (1) it provides a systematic and easy method to evaluate the cumulants of FCS by avoiding the tedious 
application of Wick's theorem; (2) more importantly, the derivation makes no assumptions about interaction between electrons 
inside the central region; and (3) we can directly use the whole power of the Feynman diagram technique and connect to many known 
results of the NGF. Of course, a proper knowledge of the self energy is still indispensable for interacting systems. With respect 
to this consideration, Gogolin and his coworkers first applied their approach to investigate the FCS of 
charge\cite{Gogolin,Komnik} current through a QD in the Kondo regime at the Toulouse limit, where the single-impurity Anderson 
Hamiltonian can be mapped to a quadratic form by performing a canonical transformation. Moreover they analyzed the FCS of both 
charge\cite{Gogolin2} and spin\cite{Schmidt} currents at the strong-coupling fixed point within the framework of the 
Nozi\`eres-Fermi-liquid theory. Besides, they also studied the current cross-correlation (CC) correlations of a multiterminal 
Kondo-QD in the strong coupling limit.\cite{Schmidt2} Another application of this approach is to explore the FCS of a molecular 
QD with strong electron-phonon interaction.\cite{Schmidt3}

However, it is known that the theories of the Toulous limit and the strong-coupling limit are both valid only at the deep Kondo 
region, where charge fluctuation is totally quenched and the dynamic properties of the system are completely determined by spin 
fluctuation. So in the present paper we will perform a complementary investigation of the FCS for a Kondo-QD in the multi-probe 
case at the deep and intermediate Kondo regions by means of the Coleman's infinite-$U$ SBMFT,\cite{Coleman} which is believed to 
provide a proper description of the Kondo correlation at these regions even under nonequilibrium situation at zero 
temperature.\cite{Lopez1,Sanchez,Lopez2} 
In combination of the SBMFT with the Gogolin and Komnik's approach, we find that the self-consistent equations, which determine 
the expectation values of the slave-boson fields, becomes counting field $\lambda$-dependent. It is argued that these 
$\lambda$-dependent terms of the mean values describe the Bose field fluctuation (it is equivalent to the charge fluctuation due 
to the completeness relation between fermions and bosons) induced by transport measurement. More interestingly, we find that 
these $\lambda$-related parts of slave-boson fields generate an additional contribution in the zero-frequency shot noise formula, 
in contrast to our previous result without consideration of the Bose field fluctuations.\cite{Dong2} Numerical calculations show 
that these additional term results in obvious enhancement of the current auto-correlation in the two-terminal setup and even a 
sign change of the current cross-correlation in the three-terminal setup of the Kondo dot in the intermediate Kondo regions.

The outline of the paper is as follows. In Sec.~II, we give the model Hamiltonian describing electron tunneling through an 
interacting QD attached to three leads in the presence of the three respective counting fields and explain the theoretical 
formulation of the SBFMT, the counting-field-dependent self-consistent equations. In Sec.~III, we provide the explicit 
expressions of the first-order differential equation of the adiabatic potential with respect to the counting fields, and the 
zero-frequency current auto- and cross-correlations at zero temperature. In particular, we discuss the additional terms due to 
fluctuation of slave-boson field. In Sec.~IV, we perform concrete numerical calculations and discussions for the current auto- 
and cross-correlations based on the formulae given in Sec.~III. Finally, our conclusions are given in section~V.

\section{Model Hamiltonian and Theoretical Formulation}

\subsection{Model}

We model the electronic transport through a single-level QD coupled to multi-terminal using the infinite-$U$ Anderson Hamiltonian 
as: 
\bn H &=& \sum_{\eta k\sigma} \epsilon_{\eta k} c_{\eta k\sigma}^{\dagger}
c_{\eta k\sigma} + \sum_\sigma \epsilon_d d_\sigma^\dagger d_\sigma + U n_{d\uparrow} n_{d\downarrow} \cr && + \sum_{\eta 
k\sigma}\left [ \gamma_\eta e^{i\lambda_\eta(t)/4} d_\sigma^\dagger c_{\eta k \sigma}+ {\rm H.c.} \right ], 
\en
where $c_{\eta k \sigma}^{\dagger}$ ($c_{\eta k \sigma }$) creates (destroys) an electron with momentum $k$, spin 
$\sigma=\{\uparrow,\downarrow\}$, and energy dispersion $\epsilon_{\eta k \sigma }$ in the lead $\eta$, with $\eta=\{1,2,3\}$ in 
the three-terminal case; $d_\sigma^\dagger$ ($d_\sigma$) creates (destroys) a spin-$\sigma$ electron on the QD with energy level 
$\epsilon_{d}$; $\gamma_\eta$ is the coupling matrix element between the dot and lead $\eta$; and $\lambda_\eta (t)$ is the 
artificially introduced measuring field with respect to the lead $\eta$ on the Keldysh contour: $\lambda_\eta(t)=\lambda_{\eta -} 
\theta(t) \theta({\cal T}-t)$ on the forward path and $\lambda_\eta(t)=\lambda_{\eta +} \theta(t) \theta({\cal T}-t)$ on the 
backward path (${\cal T}$ is the measuring time during which the counting fields are non-zero and $\lambda_{\eta 
-}=-\lambda_{\eta +}=\lambda$).\cite{Levitov1,Gogolin,Komnik}

Under the framework of the infinite-$U$ slave-boson approach, the ordinary electron operators on the QD are decomposed into a 
boson operator $\hat b$ (describing the empty state on the QD) and a fermion operator $f_\sigma$ (denoting the singly occupied 
state with electron spin-$\sigma$), $d_\sigma= f_\sigma \hat b^\dagger$ and $d_\sigma^\dagger = f_\sigma^\dagger \hat 
b$.\cite{Coleman} In addition, a constraint, $\hat b^\dagger \hat b + \sum_\sigma f_\sigma^\dagger f_\sigma=1$, must be imposed 
on these auxiliary operators as requirement of no double occupancy in the $U\rightarrow\infty$ limit. Then, the effective 
Hamiltonian becomes
\begin{eqnarray}
H_{eff} &=& \sum_{\eta k\sigma}\epsilon_{\eta k}c^{\dagger}_{\eta k\sigma}c_{\eta k\sigma} + \sum_\sigma
\epsilon_d f_\sigma^\dagger f_\sigma \cr
&& + \sum_{\eta k\sigma} \left [ \gamma_\eta e^{i\lambda_\eta(t)/4} f_\sigma^\dagger \hat b c_{\eta k \sigma} + {\rm H.c.} \right 
] \cr
&& + \xi \left ( \hat b^\dagger \hat b + \sum_\sigma f_\sigma^\dagger f_\sigma -1 \right ) , \label{ham}
\end{eqnarray}
with a Lagrange multiplier $\xi$ to guarantee satisfaction of the constraint.

\subsection{Self-consistent Equations of Slave-Boson Mean Field Theory}

In the mean field approximation, the slave Bose operators, $\hat b^\dagger$ and $\hat b$, can be assumed as $c$-number and 
replaced by their corresponding expectation values, $b^\dagger$ and $b$, in the Hamiltonian Eq.\,(\ref{ham}). To determine the 
unknown parameters, $b$ and $\xi$, we start from the constraint, and the equation of motion of the slave-boson operator using the 
Keldysh technique for systems out of equilibrium:
\bn
\sum_{\sigma} G_{f\sigma}^{-+}(t,t) + n_b &=& 1, \\
\xi b + \sum_{\eta k \sigma} \gamma_\eta G_{f\sigma,\eta k \sigma}^{-+}(t,t) &=& 0,
\en
where $n_b=\langle \hat b^\dagger \hat b \rangle$ is the unoccupied-state number, $G_{f\sigma}^{-+}(t,t') = i\langle 
f_\sigma^\dagger (t') f_\sigma (t) \rangle$ is the dot non-equilibrium correlation GF, and $G_{f\sigma,\eta k 
\sigma}^{-+}(t,t')=i\langle c_{\eta k \sigma}^\dagger (t') f_\sigma(t) \rangle$ denotes the dot-lead correlation GF. With the 
Hamiltonian Eq.\,(\ref{ham}), the mixed correlation GF can be readily cast in terms of $G_{f\sigma}^{-+}(t,t')$ with the help of 
the equation of motion of the operators and then applying the Langreth analytical continuation rules in a complex time 
contour.\cite{Langreth} In this end, we obtain a self-consistent set of equations in terms of the QD's correlation GF in Fourier 
space:
\bn
\frac{1}{2\pi i} \sum_{\sigma} \int d\omega \, G_{f\sigma}^{-+}(\omega) + n_b &=& 1, \label{sceq1} \\
\xi + \frac{1}{2\pi i} \sum_{\sigma} \int d\omega\, G_{f\sigma}^{-+} (\omega) (\omega-\epsilon_d-\xi) &=& 0. \label{sceq2}
\en

Therefore, our next step is to calculate the dot correlation GF, $G_{f\sigma}^{-+}(\omega)$. It is evident that the mean-field 
Hamiltonian Eq.\,(\ref{ham}) is effectively a noninteracting resonant level model. By taking the counting field 
$\lambda_{\eta}(t)$ to be opposite constant on the forward and backward Keldysh branches as $\lambda_{\eta-}=-\lambda_{\eta 
+}=\lambda_\eta$,\cite{Levitov1} one can therefore readily evaluate the NGF of pseudo-fermion operator $f_\sigma$ in terms of the 
original notation of Keldysh for GFs (the time-ordered GFs):
\begin{widetext}
\bq
{\bm G}_{f\sigma}(\omega) = {1\over {\cal D}_0(\omega)} \left (
\begin{array}{cc}
\omega-\epsilon_d- \xi + i n_b \sum_\eta \Gamma_\eta (2f_\eta-1) & 2i n_b \sum_\eta e^{i\lambda_\eta/2} \Gamma_\eta f_\eta \\
-2i n_b \sum_\eta e^{-i\lambda_\eta/2} \Gamma_\eta(1- f_\eta) & -(\omega-\epsilon_d-\xi) + i n_b \sum_\eta \Gamma_\eta 
(2f_\eta-1) \\
\end{array}
\right ), \label{gf}
\eq
with ($\Gamma=\sum_\eta \Gamma_\eta$)
\bq
{\cal D}_0(\omega) = (\omega-\epsilon_d-\xi)^2+(n_b\Gamma)^2 + 4 n_b^2 \sum_{\eta,\eta'} \Gamma_\eta \Gamma_{\eta'} f_\eta 
(1-f_{\eta'}) \left ( e^{i(\lambda_\eta - \lambda_{\eta'})/2}-1 \right ) ,
\eq
\end{widetext}
where $f_\eta =[1+\exp {(\omega-\mu_\eta)/k_{\rm B}T}]^{-1}$ is the Fermi distribution function at temperature $T$ and chemical 
potential $\mu_{\eta}=E_F+eV_{\eta}$ of lead $\eta$ ($E_F$ is the Fermi energy and $V_{\eta}$ is the bias-voltage applied to lead 
$\eta$), and $\Gamma_\eta = \pi \sum_{k} |V_\eta|^2 \delta(\omega-\epsilon_{\eta k \sigma})$ is the coupling strength between the 
QD and lead $\eta$.  In the wide band limit, we neglect the energy dependence of $\Gamma_{\eta}$ and take it as a constant. Note 
that the GF formulae is similar to that of a noninteracting system, except with the effective energy level $\epsilon_d+\xi$ and 
the effective tunnel-coupling constant $n_b \Gamma_\eta$ instead, which renormalize the GF of the QD due to Kondo correlation 
under the approximation employed here.

Substituting the resulting GF Eq.\,(\ref{gf}) into the self-consistent equations (\ref{sceq1}) and (\ref{sceq2}), we can obtain 
the two parameters, $n_b$ and $\xi$, under a finite bias voltage. Obviously, they are both functions of the counting fields
$\lambda_\eta$. For the sake of analysis and calculation of the shot noise, we can expand $n_b$ and $\xi$ to the first order of 
$\lambda_\eta$ as:
\bn
n_b &=& n_b^{(0)} + i\sum_\eta \frac{1}{2} \lambda_\eta n_{b\eta}^{(1)} + {\cal O}(\lambda_\eta^2), \\
\xi &=& \xi^{(0)} + i\sum_\eta \frac{1}{2} \lambda_\eta \xi^{(1)}_\eta + {\cal O}(\lambda_\eta^2),
\en
where the superscripts $(0)$ and $(1)$ denotes the zeroth and first order terms of the expansion coefficients, respectively. The 
zeroth order terms, $n_b^{(0)}$ and $\xi^{(0)}$, are the original variational parameters of the SBMFT and irrespective of the 
counting fields; while the first order terms, $n_{b\eta}^{(1)}$ and $\xi_\eta^{(1)}$, are new here depicting fluctuations of the 
two parameters around their respective expectation values due to measurement.

The zeroth order self-consistent equations become
\bn
n_b^{(0)} \left [ \int {d\omega\over {\pi}} {2\sum_\eta \Gamma_\eta f_\eta \over {(\omega- \tilde\epsilon_d)^2 + \tilde\Gamma^2}} 
+ 1 \right ] &=& 1, \label{seq0-1} \\
\xi^{(0)} + \int {d\omega\over {\pi}}{2 (\omega-\tilde\epsilon_d){(\sum_\eta \Gamma_\eta f_\eta)}\over 
{(\omega-\tilde\epsilon_d)^2 + \tilde\Gamma^2}} &=& 0, \label{seq0-2}
\en
with $\tilde\epsilon_d=\epsilon_d+\xi^{(0)}$ and $\tilde\Gamma=n_b^{(0)}\Gamma$. It is noted that these two nonlinear equations 
are exactly the same as the previous results without the presence of the counting fields $\lambda_{\eta}$,\cite{Sanchez,Coleman} 
and they constitute a closed set of equations to completely determine the unknown parameters, $n_b^{(0)}$ and $\xi^{(0)}$, for a 
particular QD system, $\epsilon_d$, at a given bias voltage. The parameter $\xi^{(0)}$ characterizes the location of the Kondo 
peak in the quasiparticle density of states, and the parameter $n_b^{(0)}$ mimics the width of the Kondo peak (i.e. the Kondo 
temperature $T_{K}$) and renormalizes the tunnel-coupling of the QD to the external leads by Kondo correlation. It is known that 
the two parameters give the correct qualitative behavior of Kondo physics at zero temperature and low bias voltages ($eV \sim 
T_{K}$), and thus define the current-bias voltage characteristics and the differential conductance of a 
Kondo-QD.\cite{Dong2,Lopez2,Sanchez,Coleman}
Furthermore, we can obtain the first order self-consistent equations as
\bn
({\cal I}^{(0)}- {\cal I}^{(1)}_a) {n_{b\eta}^{(1)} \over  n_b^{(0)}} + {\cal I}^{(1)}_b \xi_\eta^{(1)} + {\cal I}^{(1)}_{c\eta} 
&=& 0, \label{seq1} \\
{\cal K}_a^{(1)} {n_{b\eta}^{(1)} \over n_b^{(0)}} - {\cal K}^{(1)}_b \xi_\eta^{(1)} + {\cal K}^{(1)}_{c\eta} &=& 0, \label{seq2}
\en
where the integrals are as follows:
\begin{subequations}
\label{rateq1}
\begin{equation}
{\cal I}^{(0)}=\int {d\omega\over{2\pi}}{2\sum_\eta\Gamma_\eta f_\eta \over {(\omega-\tilde\epsilon_d)^2 + \tilde\Gamma^2}},
\end{equation}
\begin{equation}
{\cal I}^{(1)}_a = \int {d\omega\over{2\pi}}{4\tilde\Gamma^2(\sum_\eta\Gamma_\eta f_\eta) \over {[(\omega-\tilde\epsilon_d)^2 + 
\tilde\Gamma^2]^2}},
\end{equation}
\begin{equation}
{\cal I}^{(1)}_b = \int {d\omega\over{2\pi}}{4(\omega-\tilde\epsilon_d) (\sum_\eta \Gamma_\eta f_\eta) \over 
{[(\omega-\tilde\epsilon_d)^2 + \tilde\Gamma^2]^2}},
\end{equation}
\bn
{\cal I}_{c\eta}^{(1)} &=& \int {d\omega\over{2\pi}} \left \{ {2\Gamma_\eta f_\eta \over {(\omega-\tilde\epsilon_d)^2 + 
\tilde\Gamma^2}} \right.  \cr
&& \left. - {8(n_b^{(0)})^2 (\sum_{\eta''}\Gamma_{\eta''} f_{\eta''}) [\sum_{\eta'} \Gamma_\eta \Gamma_{\eta'} (f_\eta-f_{\eta'}) 
] \over {[(\omega-\tilde\epsilon_d)^2 + \tilde\Gamma^2 ]^2}} \right \} , \cr
&&
\en
\begin{equation}
{\cal K}^{(1)}_a = 2\tilde\Gamma^2 {\cal I}^{(1)}_b,
\end{equation}
\begin{equation}
{\cal K}^{(1)}_b = 1 + \int {d\omega\over{2\pi}} {4 [ (\omega-\tilde\epsilon_d)^2 - \tilde\Gamma^2 ] (\sum_\eta \Gamma_\eta 
f_\eta) \over {[(\omega-\tilde\epsilon_d)^2 + \tilde\Gamma^2]^2}},
\end{equation}
\bn
{\cal K}_{c\eta}^{(1)} &=& \int {d\omega \over {2\pi}} 16(n_b^{(0)})^2 (\omega-\tilde\epsilon_d) ( \sum_{\eta''} \Gamma_{\eta''} 
f_{\eta''} ) \cr
&& \times {\sum_{\eta'} \Gamma_\eta \Gamma_{\eta'} (f_\eta - f_{\eta'}) \over {[(\omega-\tilde\epsilon_d)^2 + 
\tilde\Gamma^2]^2}}.
\en
\end{subequations}
These equations are one of the central results of this paper. Once $n_b^{(0)}$ and $\xi^{(0)}$ are known, $n_{b\eta}^{(1)}$ and 
$\xi_\eta^{(1)}$ can be evaluated by solving the set of linear equations, Eqs.\,(\ref{seq1}) and (\ref{seq2}). The two first 
order terms have no influence on the current but do affect the shot noise and of course the higher cumulants (see below). One can 
argue that the two terms depict the fluctuations of both the boson field $\hat b$ and the renormalization of the resonant peak 
$\xi$, which can be ascribed to the charge fluctuation of the QD due to its attachment to external electrodes and application of 
bias voltage. Consequently, we will find in the following calculation that these terms have a nonzero contribution to the shot 
noise of the QD at a finite bias voltage. Moreover, they play a more important role at the intermediate Kondo regions, where the 
charge fluctuation is more profound, than at the deep Kondo regions.

\section{Full Counting Statistics By Slave-Boson Mean Field Theory}

In this section, we will first study the cumulant generating function (CGF) $\chi(\lambda)\equiv 
\chi(\lambda_1,\lambda_2,\lambda_3)$ for the FCS of a Kondo-QD with three terminals within SBMFT. Then we will derive the formula 
for the current auto-correlation and cross-correlation taking into account the charge fluctuation effect in the two- and 
three-terminal configurations, respectively. 

The CGF can be calculated as a Keldysh partition function\cite{Levitov1}
\bq
\chi(\lambda)= \langle T_C e^{-i \int_C T_\lambda(t) dt} \rangle,
\eq
where $T_C$ is the Keldysh contour ordering operator and $T_\lambda(t)=\sum_{\eta k\sigma} \left [ \gamma_\eta 
e^{i\lambda_\eta(t)/4} f_\sigma^\dagger \hat b c_{\eta k \sigma} + {\rm H.c.} \right ]$ is the electron tunneling operator in the 
Hamiltonian Eq.~(\ref{ham}). 
According to Refs.~\onlinecite{Gogolin,Komnik}, to calculate the CGF $\chi(\lambda)$ it is technically more convenient employing 
the {\em adiabatic potential} method: $\ln \chi(\lambda)=-i{\cal T } {\cal U}(\lambda_-, \lambda_+)=-i{\cal T } {\cal U}(\lambda, 
-\lambda)$, where the adiabatic potential ${\cal U}(\lambda_-, \lambda_+)$ is defined due to the nonequilibrium Feynman-Hellmann 
theorem as
\begin{eqnarray}
{\partial {\cal U}(\lambda_-, \lambda_+)\over {\partial \lambda_{\eta -}}} &=& \Big\langle {\partial T_\lambda(t)\over{\partial 
\lambda_{\eta -}}} \Big\rangle_\lambda \nonumber\\
&=& {i\over 4}\sum_{k\sigma} \langle \gamma_\eta e^{i\lambda_{\eta -}/4} f_\sigma^\dagger \hat b c_{\eta k \sigma} - {\rm H.c.} 
\rangle_\lambda. \cr
&& \label{adp0}
\end{eqnarray}
Therefore, the evaluation of the adiabatic potential amounts to a calculation of the mixed GFs, which can be cast into a 
combination of the dot GF ${\bm G}_{f\sigma}(\omega)$ and bare lead GFs. 
By using the NGF obtained in last section and taking the counting field $\lambda_{\eta -}=-\lambda_{\eta+}=\lambda_\eta$, the 
derivative of the adiabatic potential is given by
\begin{widetext}
\bn 
{\partial {\cal U}(\lambda, -\lambda)\over {\partial
\lambda_\eta}}=-2\int {d\omega\over{2\pi}} {\sum_{\eta'} T_{\eta\eta'}(\omega) [f_\eta (1-f_{\eta'}) 
e^{i(\lambda_\eta-\lambda_{\eta'})/2} - f_{\eta'} (1-f_\eta) e^{-i(\lambda_\eta-\lambda_{\eta'})/2}] \over {1 + \sum_{\eta,\eta'} 
T_{\eta\eta'}(\omega) f_\eta (1-f_{\eta'})
[ e^{i(\lambda_\eta-\lambda_{\eta'})/2}-1]}}, \label{adp} 
\en
\end{widetext}
where $T_{\eta\eta'}(\omega)$ is the transmission coefficient of electron between leads $\eta$ and
$\eta'$:
\begin{equation}
T_{\eta\eta'}(\omega) = {4n_b^2\Gamma_\eta\Gamma_{\eta'} \over {(\omega-\epsilon_d-\xi)^2 + (n_b\Gamma)^2}}.
\end{equation}
It is clear to see that the transmission coefficient (1) has a simple Lorentzian line shape centered around the Fermi level (the 
Abrikosov-Suhl resonance) with a renormalized width due to Kondo correlation, and (2) is symmetric for exchanging $\eta$ and 
$\eta'$: $T_{\eta\eta'}(\omega)=T_{\eta'\eta}(\omega)$. It should be noted that due to the counting field dependence of the 
parameters $n_b$ and $\xi$ involved in the transmission coefficient $T_{\eta\eta'}(\omega)$, an explicit analytic expression for 
the CGF cannot be recovered here from Eq.~(\ref{adp}). However, one can still obtain the $m$-th cumulant of the charge 
distribution involving lead $\eta$ by executing the $m$-th derivative of the CGF, with respect to the corresponding counting 
field $\lambda_\eta$ at $\lambda_1=\lambda_2=\lambda_3=0$ (we denote this condition as $\lambda=0$ for shorthand in the 
following). In particular, the first two cumulants give the average current and the zero-frequency shot noise, respectively, 
which constitute our main objects in the present paper.
For example, the current $I_\eta$ from lead $\eta$ to the QD is evaluated as follows:
\begin{widetext}
\begin{eqnarray}
I_\eta &=& \sum_\sigma I_{\eta\sigma} = {2e\over \hbar} {1 \over {\cal T}} {\partial \ln \chi(\lambda)\over {\partial 
(i\lambda_\eta/2)}} {\bigg |}_{\lambda=0} = -{4e\over \hbar} {\partial {\cal U}(\lambda, -\lambda)\over {\partial
\lambda_\eta}}{\bigg |}_{\lambda=0} \cr
&=& {2e\over \hbar} \int {d\omega\over{2\pi}} {\sum_{\eta'} T_{\eta\eta'}(\omega) [f_\eta (1-f_{\eta'}) 
e^{i(\lambda_\eta-\lambda_{\eta'})/2} - f_{\eta'} (1-f_\eta) e^{-i(\lambda_\eta-\lambda_{\eta'})/2}] \over {1 + \sum_{\eta,\eta'} 
T_{\eta\eta'}(\omega) f_\eta (1-f_{\eta'}) (e^{i(\lambda_\eta-\lambda_{\eta'})/2}-1)}} {\bigg |}_{\lambda=0} \cr
&=& {2e\over h}\int d\omega \sum_{\eta'} T_{\eta\eta'}(\omega) (f_\eta-f_{\eta'}) \mid_{\lambda=0}. \label{current}
\end{eqnarray}
It is clear that the three-terminal current formula Eq.\,(\ref{current}) is in perfect agreement with the previous 
result,\cite{Dong2,Sanchez} and reduces to the transparent form of the famous Meir-Wingreen current formula in the two-terminal 
case (for instance, by setting $\Gamma_3=0$).\cite{Meir2}

We calculate now the zero-frequency current fluctuations of lead $\eta$:
\begin{eqnarray}
S_{\eta\eta}(0) = {4e^2\over \hbar}{1 \over {\cal T}}{\partial^2 \ln \chi(\lambda) \over {\partial (i\lambda_\eta/2)^2}} {\bigg 
|}_{\lambda=0} = S_{\eta\eta}^{(0)}(0)+S_{\eta\eta}^{(c)}(0),
\end{eqnarray}
where $S_{\eta\eta}^{(0)}(0)$ is the mean-field result of the current correlation
\bn
S_{\eta\eta}^{(0)}(0) = {4e^2\over h} \int d\omega \left \{ \sum_{\eta'(\neq\eta)} T_{\eta\eta'}(\omega) [ f_\eta (1-f_{\eta'}) 
\right.
+ f_{\eta'} (1-f_\eta) ]
 \left. - \left [ \sum_{\eta'} T_{\eta{\eta'}}(\omega) (f_\eta-f_{\eta'}) \right ]^2 \right \}{\bigg |}_{\lambda=0}, 
\label{automf}
\en
and $S_{\eta\eta}^{(c)}(0)$ is the correction to the mean-field result of the shot noise
\begin{equation}
S_{\eta\eta}^{(c)}(0) = {4e^2 \over h} \int d \omega \sum_{\eta'} \left [ {\partial T_{\eta\eta'}(\omega) \over {\partial 
(i\lambda_{\eta}/2)}} {\bigg |}_{\lambda=0} (f_\eta-f_{\eta'}) \right ].
\end{equation}
It is clear that $S_{\eta\eta}^{(0)}(0)$ depends on the zeroth order terms of the two parameters, $n^{(0)}$ and $\xi^{(0)}$ only, 
while $S_{\eta\eta}^{(c)}(0)$ is relevant to the first order terms, $n_{b\eta}^{(1)}$ and $\xi_{\eta}^{(1)}$, and can be 
explicitly expressed as
\bn
S_{\eta\eta}^{(c)}(0) = {4e^2\over h} \int d\omega \sum_{\eta'} \left \{ {8n_{b\eta}^{(1)} \over {n^{(0)}_b}} 
{(\omega-\tilde\epsilon_d)^2 \tilde\Gamma_{\eta} \tilde\Gamma_{\eta'} \over {\left[ (\omega -\tilde\epsilon_d)^2 + \tilde\Gamma^2 
\right ]^2 }} \right.
\left. + 8 \xi_{\eta}^{(1)} { (\omega - \tilde\epsilon_d) \tilde\Gamma_{\eta} \tilde\Gamma_{\eta'} \over 
{(\omega-\tilde\epsilon_d)^2 + \tilde\Gamma^2 }} \right \} (f_\eta - f_{\eta'}). \label{autocor}
\en
The zero-frequency CC of currents through the different leads $\eta\neq\eta'$ can be calculated as
\bq
S_{\eta\eta'}(0) = {4e^2 \over \hbar} {1\over {\cal T}} {\partial^2 \ln \chi(\lambda) \over {\partial (i\lambda_\eta/2) \partial 
(i\lambda_{\eta'}/2)}} {\bigg |}_{\lambda=0} = S_{\eta\eta'}^{(0)}(0) + S_{\eta\eta'}^{(c)}(0).
\eq
Likewise, the CC $S_{\eta\eta'}(0)$ can be separated into two parts, the mean field result $S_{\eta\eta'}^{(0)}(0)$ and its 
correction $S_{\eta\eta'}^{(c)}(0)$:
\bq
S_{\eta\eta'}^{(0)}(0) = -{4e^2 \over h} \int d\omega \left \{ T_{\eta\eta'}(\omega) [f_\eta (1-f_{\eta'}) + f_{\eta'} (1-f_\eta) 
] + \left [ \sum_{\eta_1} T_{\eta{\eta_1}}(\omega) (f_\eta - f_{\eta_1}) \right ]
 \left [ \sum_{\eta_2} T_{\eta'{\eta_2}}(\omega) (f_{\eta'}-f_{\eta_2}) \right ] \right \} {\bigg |}_{\lambda=0} , 
\label{crossmf}
\eq
\bn
S_{\eta\eta'}^{(c)}(0) &=& {4e^2\over h} \int d\omega \sum_{\eta_1} \left [ {\partial T_{\eta\eta_1}(\omega)\over {\partial 
(i\lambda_{\eta'}/2)}} {\bigg |}_{\lambda=0} (f_\eta-f_{\eta_1}) \right ] \cr
&=& {4e^2\over h} \int d\omega \sum_{\eta_1} \left \{ { 8 n^{(1)}_{b\eta'} \over {n^{(0)}_b}} { (\omega-\tilde\epsilon_d)^2 
\tilde\Gamma_\eta \tilde\Gamma_{\eta_1} \over {\left [ (\omega-\tilde\epsilon_d)^2 + (\tilde\Gamma)^2 \right ]^2} } \right. 
\left. + 8 \xi^{(1)}_{\eta'} { (\omega-\tilde\epsilon_d) \tilde\Gamma_\eta \tilde\Gamma_{\eta_1} \over { ( \omega - 
\tilde\epsilon_d )^2 + (\tilde\Gamma)^2 }} \right \} (f_\eta-f_{\eta_1}). \label{crosscor}
\en
For illustration, we give the explicit expression for the mean-field shot noise, i.e., the current auto-correlation measured at 
the same lead (e.g., lead $1$):
\begin{eqnarray}
S_{11}^{(0)}(0) &=& {4e^2\over h}\int d\omega \{T_{12}[f_1 (1 - f_1) + f_2 (1-f_2)] + T_{13}[f_1 (1-f_1) + f_3 (1-f_3)] 
\nonumber\\
&& + T_{12} (1- T_{12}) (f_1-f_2)^2 + T_{13} (1-T_{13}) (f_1 - f_3)^2 - 2 T_{12} T_{13} (f_1-f_2) (f_1-f_3) \}\mid_{\lambda=0}, 
\end{eqnarray}
and the explicit expression for the mean-field current CC between leads $2$ and $3$:
\begin{eqnarray}
S_{23}^{(0)}(0) &=& -{4e^2\over h} \int d\omega [ T_{23} [f_2 (1 - f_2) + f_3 (1 - f_3)] + T_{23} (1-T_{23}) (f_2-f_3)^2 + T_{21} 
T_{31} (f_2-f_1)(f_3-f_1) \cr
&& + T_{21}T_{32} (f_2-f_1)(f_3-f_2) + T_{31}T_{23} (f_3-f_1)(f_2-f_3)] \mid_{\lambda=0}.
\end{eqnarray}
\end{widetext}

Furthermore, in order to compare the present results with the previous formulae, we discuss the symmetrized shot noise in the 
case of a QD connected to two reservoirs. By setting $\Gamma_3=0$, the model employed above becomes a two-terminal system and the 
symmetrized shot noise is then defined as $S(0)=[S_{11}(0)+S_{22}(0)-2S_{12}(0)]/4=S^{(0)}(0)+S^{(c)}(0)$, whose first part can 
be calculated via Eqs.\,(\ref{automf}) and (\ref{crossmf}) as
\begin{eqnarray}
S^{(0)}(0) &=& {4e^2\over h}\int d\omega \{T_{12}[f_1 (1 - f_2) + f_2 (1-f_1)] \nonumber\\
&& - T_{12}^2 (f_1-f_2)^2 \}\mid_{\lambda=0}. \label{snmf}
\end{eqnarray}
Obviously, this formula is exactly the same as our previous derivation applying Wick theorem within the framework of the 
SBFMT.\cite{Dong2} This is the reason that we call it and Eqs.\,(\ref{automf}) and (\ref{crossmf}) as {\em mean-field} terms of 
the current correlations. For reference, we briefly recall our previous derivation. The current operator $\hat I_{\eta}$ flowing 
from the lead $\eta$ to the QD is defined as
\begin{eqnarray}
\hat{I}_\eta(t) &=& -i\frac{e}{\hbar} \Big [H_{\rm eff}, \sum_{\eta k \sigma} c_{\eta k \sigma}^{\dagger} c_{\eta k \sigma} \Big 
] \cr
&=& i\frac{e}{\hbar} \sum_{\eta k \sigma} (V_\eta b^\dagger c_{\eta k \sigma}^\dagger f_{\sigma} - V_\eta^{*} f_{\sigma}^\dagger 
b c_{\eta k \sigma}), \label{i}
\end{eqnarray}
and the power spectrum of the CC correlator reads
\bq
S_{\eta\eta'}(\omega) = 2 \int d\tau e^{i\omega \tau} [ \langle \{ \hat I_{\eta}(t), \hat I_{\eta'}(t') \}\rangle - \langle \hat 
I_{\eta} \rangle \langle \hat I_{\eta'} \rangle ], \label{sn}
\eq
with $\tau=t-t'$. Calculations of the correlator involves a pair of creation and destruction operators for each of the current 
operators at the two time variables. According to the conventional diagrammatic expansion technique of GF, the statistical 
expectation of the product of any two operators $\hat{\cal A}$ and $\hat{\cal B}$ can be divided into two parts
\begin{equation}
\langle \hat{\cal A}\hat{\cal B}\rangle=\langle\hat{\cal A}\rangle\langle\hat{\cal B}\rangle+\langle\hat{\cal A}\hat{\cal 
B}\rangle_{c}, \label{wick}
\end{equation}
where $\langle\cdots \rangle_{c}$ means the connected part of the statistical average, which is vanishing for noninteracting 
systems. As a result, under the framework of SBMFT and neglecting the fluctuations of both $n_b$ and $\xi$ (i.e. only considering 
the leading terms, $n_b^{(0)}$ and $\xi^{(0)}$), we can eliminate the connected part of the above statistical expectation, and 
entirely contract the ensemble averaging with a product of four operators in $\langle \{ \hat I_{\eta}(t), \hat I_{\eta'}(t') 
\}\rangle$ as a product of ensemble averagings of two pairs of operators, such as
\bn
&& \hspace{-1cm}\langle c_{\eta k \sigma}^{\dagger}(t) f_{\sigma}(t) c_{\eta' k \sigma}^{\dagger}(t') f_{\sigma}(t') \rangle \cr
&=& \langle c_{\eta k \sigma}^{\dagger}(t) f_{\sigma}(t) \rangle \langle c_{\eta' k \sigma}^{\dagger}(t') f_{\sigma}(t') \rangle 
\cr
&& + \langle c_{\eta k \sigma}^{\dagger}(t) f_{\sigma}(t') \rangle \langle f_{\sigma}(t) c_{\eta' k \sigma}^{\dagger}(t') 
\rangle,
\en
and
\bn
&& \hspace{-1cm}\langle c_{\eta k \sigma}^{\dagger}(t) f_{\sigma}(t) f_{\sigma}^\dagger(t') c_{\eta' k \sigma}(t') \rangle \cr
&=& \langle c_{\eta k \sigma}^{\dagger}(t) f_{\sigma}(t) \rangle \langle f_{\sigma}^\dagger(t') c_{\eta' k \sigma}^{\dagger}(t') 
\rangle \cr
&& + \langle c_{\eta k \sigma}^{\dagger}(t) c_{\eta' k \sigma}(t') \rangle \langle f_{\sigma}(t) f_{\sigma}^\dagger(t')\rangle.
\en
All first terms in the left-hand side cancel out the term $\langle \hat I_{\eta} \rangle \langle \hat I_{\eta'} \rangle$ of 
Eq.\,(\ref{sn}), whereas the second terms result in two lesser GFs running antiparallel in time (one starts at $t$ going to $t'$, 
the other goes from $t'$ to $t$), according to the Langreth theorem: the object with two antiparallel GFs ${\cal C}(t,t')={\cal 
A}(t',t){\cal B}(t,t')$ has the lesser component of ${\cal C}^<(t,t')={\cal A}^<(t',t) {\cal B}^>(t,t')$. In succession, the 
hybird lesser GFs can further simplified as a proper combination of the local Keldysh GFs of the QD with the tunnel-coupling and 
Fermi functions of the leads, which leads in the end to the mean-field formula of the shot noise, Eq.\,(\ref{snmf}).

Now we concentrate on the second part of the symmetrized shot noise, $S^{(c)}(0)$. Without loss of generality, we can assume 
$\lambda_1=-\lambda_2$ (as a matter of fact, in the two-terminal case, although two counting fields $\lambda_1$ and $\lambda_2$ 
are introduced, the final result depends only on the difference $\lambda\equiv\lambda_1-\lambda_2$ due to the charge 
conservation), and define $n_{b1}^{(1)}-n_{b2}^{(1)}\equiv n_{b}^{(1)}$ and $\xi_{1}^{(1)}-\xi_{2}^{(1)}\equiv\xi^{(1)}$.
Then from Eqs.\,(\ref{autocor}) and (\ref{crosscor}), we yield
\bn
S^{(c)}(0) &=& {4e^2\over h} \int d\omega \left \{ { 8 n^{(1)}_{b} \over {n^{(0)}_b}} { (\omega-\tilde\epsilon_d)^2 
\tilde\Gamma_1 \tilde\Gamma_{2} \over {\left [ (\omega-\tilde\epsilon_d)^2 + (\tilde\Gamma)^2 \right ]^2} } \right. \cr
&& \left. + 8 \xi^{(1)} { (\omega-\tilde\epsilon_d) \tilde\Gamma_1 \tilde\Gamma_{2} \over { ( \omega - \tilde\epsilon_d )^2 + 
(\tilde\Gamma)^2 }} \right \} (f_1 - f_{2}). \label{sncor}
\en
From technical point of view, this term is a direct result of the connected part in Eq.\,(\ref{wick}), and consequently 
constitutes a correction to the mean-field result, Eq.\,(\ref{snmf}). Meanwhile, bearing in mind that in above mean-field 
derivation the fluctuations of the boson field and the renormalization of the resonant level are both neglected, one can 
immediately conclude that this term $S^{(c)}(0)$ is physically an account of the bias-voltage-induced fluctuation effect of the 
boson field. It is clear to see that $S^{(c)}(0)$ is proportional to both $n_{b}^{(1)}$ and $\xi^{(1)}$, and is vanishing under 
equilibrium situation.

\section{Results and discussion}

In this section, we present the results of numerical calculations of the zero-frequency auto- and cross-current noises for the QD 
in the two- and three-terminal cases in the Kondo regime at zero temperature, paying special attention to the effect of boson 
field fluctuation on the current correlations, i.e., the correction terms of the shot noise formulae. The self-consistent 
parameters, $n_{b}^{(0)}$ and $\xi^{(0)}$, $n_{b}^{(1)}$ and $\xi^{(1)}$, can be obtained by solving Eqs.\,(\ref{seq0-1}) and 
(\ref{seq0-2}), (\ref{seq1}) and (\ref{seq2}), respectively, for each bias voltage and a given system $\epsilon_d$. Using these 
calculated parameters, we can then compute the mean-field results of shot noises and the total shot noises as functions of bias 
voltage according to Eqs.\,(\ref{crossmf}), (\ref{crosscor}), (\ref{snmf}), and (\ref{sncor}), respectively.
In particular, we choose $\Gamma_1=\Gamma_2=\Gamma/2$ as a typical example and set $\Gamma$ as the energy unit. The reference 
energy will be always set at $E_F=0$.
For the multiterminal Kondo-QD, the width of the Kondo peak at the Fermi energy is characterized by the Kondo termperature, 
$k_BT_{\rm K}=D\exp{(-\pi |\epsilon_d|/2\Gamma)}$, at equilibrium ($D$ is the energy cutoff and is set to $D=100\Gamma$ here). In 
the following calculations, we confine the bias voltage from $0$ to $3k_BT_{\rm K}/e$ ($eV\leq 3 k_BT_{\rm K}$) to ensure that 
the SBFMT is applicable for the Kondo correlation at nonequlibrium.

\begin{figure}[t]
\includegraphics[height=8cm,width=7cm]{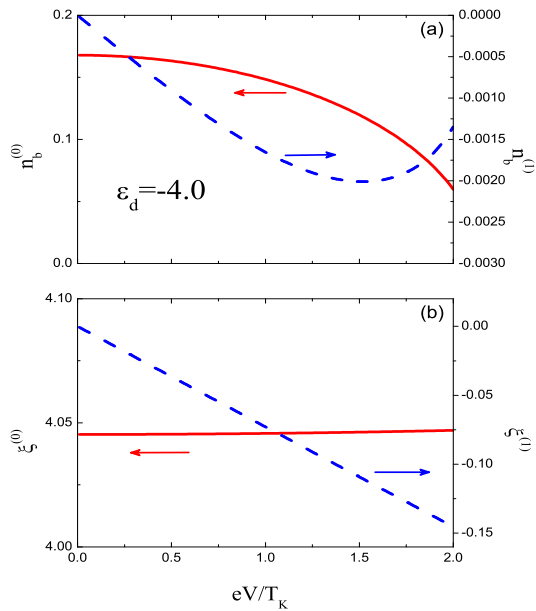}
\caption{(Color online) The calculated self-consistent parameters, $n_{b}^{(0)}$ and $n_{b}^{(1)}$ (a), $\xi^{(0)}$ and 
$\xi^{(1)}$ (b) as functions of bias voltage $V$ for the two-terminal QD with the energy level $\epsilon_d=-4.0$.}
\label{fig1}
\end{figure}

First, we discuss the auto-current correlation for the two-terminal case, in which the external bias voltage is applied 
symmetrically, $V_1=-V_2=V/2$. For illustration, we depict in Fig.~1 the calculated self-consistent parameters, the zeroth order 
terms $n_{b}^{(0)}$, $\xi^{(0)}$, and the first order terms $n_{b}^{(1)}$, $\xi^{(1)}$ versus the bias voltage $V$ for a QD with 
the discrete energy level $\epsilon_d=-4.0$. It is easy to see that (1) these first order terms are nearly two orders of 
magnitude smaller than their corresponding zeroth order terms at the whole region of the bias voltages; (2) they are increasing 
with increase of applied bias voltage; (3) they are vanishing at equilibrium, $V=0$. These features are in good agreement with 
the consideration that the first order terms describe the fluctuations of the corresponding parameters stemming from the 
bias-voltage-induced charge fluctuations even in the Kondo regime. However, these small fluctuations do cause an additional 
positive contribution to the mean-field shot noise as shown in Fig.~2(a,b), where the mean-field results $S^{(0)}(0)$ and the 
total results $S(0)$ of the symmetrized shot noise are both plotted as functions of bias voltage for two QDs with 
$\epsilon_d=-5.0$, $-4.0$. Meanwhile, it is clear that the enhancement is more obvious in the system with $\epsilon_d=-4.0$ than 
that of the system with $\epsilon_d=-5.0$. This observation is consistent with the fact that the charge fluctuation effect is 
entirely quenched in the deep Kondo regime but begins to play a unambiguous role in determining dynamic properties of the Kondo 
system when the system departs away from the deep Kondo regime. In Fig.~2(c,d), we plot the Fano factor $\gamma=S(0)/2eI$ [$I$ is 
the current calculated according to Eq.\,(\ref{current})] as well.

\begin{figure}[hbt]
\includegraphics[height=8cm,width=8.5cm]{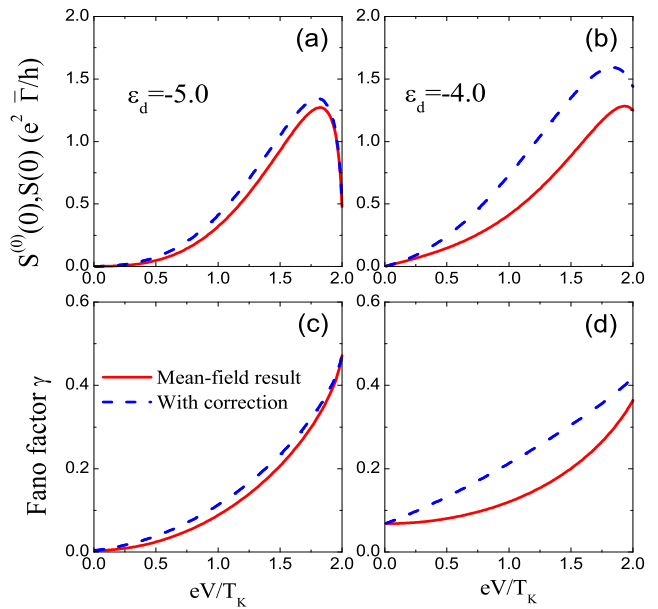}
\caption{(Color online) The symmetrized zero-frequency shot noises (a,b) and the Fano factors (c,d) as functions of bias voltage 
in the two-terminal case for the Kondo-QD with several energy levels, $\epsilon_d=-5.0$ (a,c) and $-4.0$ (b,d). The mean-field 
results and the total shot noises are plotted as solid-red lines and dash-blue lines, respectively.}
\label{fig2}
\end{figure}

We now turn to the current CC in the case of three-terminal QD for $V_1=-V_2=-V_3=V/2$ and equal tunnel-couplings 
$\Gamma_3=\Gamma/2$. In Fig.~3, we show the zero-frequency CCs with and without the contribution of the Bose field fluctuation, 
$S_{23}(0)$ and $S_{23}^{(0)}(0)$, and the corresponding Fano factors defined as $\gamma_{23}=S_{23}(0)/2e\sqrt{|I_2 I_3|}$. It 
is well-known that in a noninteracting system, the sign of the current CC between different normal-metallic leads is always 
negative (antibunching) due to the Pauli exclusion principle of Fermionic statistics of electrons,\cite{Blanter} which has been 
confirmed experimentally in a Hanbury-Brown-Twiss setup (HBT).\cite{HBT} Nevertheless, a positive current CC (bunching) has been 
predicted in certain situations: a hybrid superconductor-normal system;\cite{Anantram} spin-dependent sequential tunneling 
through a QD with three ferromagnetic leads;\cite{Cottet} a Coulomb interaction coupled system.\cite{Martin,Dong4} Very recently, 
a sign change of the current CC due to dynamical channel blockade has been experimentally observed in a subtle measurement of the 
CC between two output terminals in sequential tunneling through two capacitively coupled QDs connected to four independent 
electrodes.\cite{McClure}

\begin{figure}[htb]
\includegraphics[height=7cm,width=8.5cm]{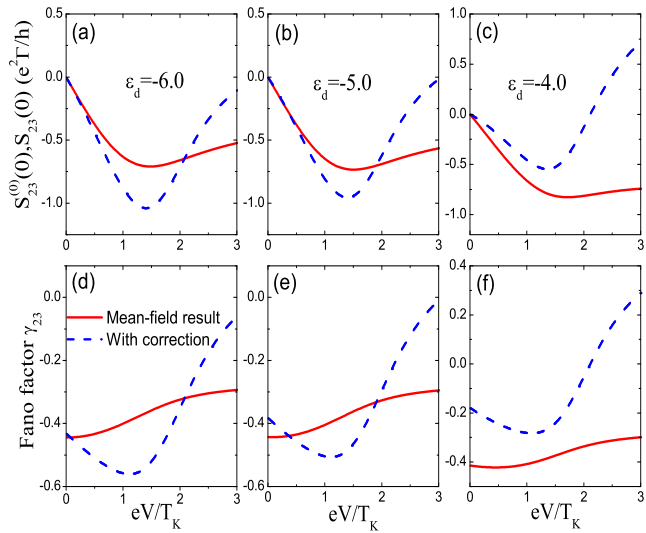}
\caption{(Color online) The current CCs between leads $2$ and $3$, $S_{23}(0)$ (a,b), and the corresponding Fano factors 
$\gamma_{23}$ (c,d) as functions of bias voltage in the injecting lead in the two-terminal case for the Kondo-QD with several 
energy levels, $\epsilon_d=-6.0$ (a,d), $-5.0$ (b,e), and $-4.0$ (c,f). The mean-field results and the total shot noises are 
plotted as solid-red lines and dash-blue lines, respectively.}
\label{fig3}
\end{figure}

For the case of Kondo-correlated transport, we notice that there have been two theoretical work devoted to the current CC in this 
regime and no sign reversal has been found.\cite{Sanchez,Schmidt2} Schmidt and his cowork have investigated the FCS of Kondo-type 
tunneling through a multi-terminal QD using the Nozi\`eres-Fermi-liquid theory and reported analytical expressions for the 
current CCs at the true strong-coupling fixed point and the departure from this point.\cite{Schmidt2}
S\'anchez and L\'opez has applied the same SBMFT as the present work to calculate the current CC of a three-terminal QD in the 
presence of ferromagnetic leads.\cite{Sanchez} Different from this paper is that they take no account of the fluctuations of both 
the Bose field and the level renormalization. In other words, their calculations correspond to the mean-field results of the 
present investigation. Therefore, our mean-field Fano factors $\gamma_{23}$ as shown in Fig.~3(d-f) by the solid-red lines has 
the same properties as those in Ref.~\onlinecite{Sanchez} in the case of paramagnetic leads: (1) it is always negative; (2) it 
has a minimum at $V=0$ and the value of the minimum is nearly equal to $-0.44$; (3) it reaches a saturation value at very high 
bias voltage.
It is however worth noticing that a rate equation calculation has addressed a sign change of the current CC due to dynamical spin 
blockade for the sequential tunneling through the same setup, i.e. a QD connected to three ferromagnetic leads.\cite{Cottet} What 
is interesting in the present calculations is that we indeed find a positive current CC for the QD with $\epsilon_d=-4.0$ under 
higher bias voltages $V> 2.0T_{\rm K}$, as shown in Fig.~3(c,f) by the dash-blue line, if the charge fluctuation is taken into 
account, even though the system under studied involves no ferromagnetic electrode. But, the current CC remains negative for the 
systems with $\epsilon_d=-6.0$ and $-5.0$, since the charge fluctuation is weaker in the two systems than that in the system with 
$\epsilon_d=-4.0$. Generally, the overall role of charge fluctuation is to reduce the fermionic HBT type correlation at higher 
bias voltage region, which is consistent with the analytical conclusion in Ref.~\onlinecite{Schmidt2}.

\section{Conclusions}

In conclusion, we have analyzed the effects of charge fluctuation on the current statistics in a Kondo QD connected to 
multi-leads at zero temperature on the basis of the infinite-$U$ SBMFT to demonstrate the Kondo correlation. By introducing 
counting fields with respect to each electrodes and employing the NGF technique, we have derived an explicit analytical 
expression for the first-order differential equation of the adiabatic potential with respect to the counting fields. Meanwhile, 
we have found that the SBMFT self-consistent equations become dependent on the counting fields, which provides a powerful 
instrument for taking account of the effect of the Bose field fluctuation, i.e. the charge fluctuation effect, at nonequilibrium 
on the current correlation. In particular, we have performed derivations for the concrete analytical expressions for the 
symmetrized shot noise in the case of two-terminal QD and the current CC in the case of three-terminal QD. It has been found that 
the consideration of the effect of charge fluctuation generates an additional contribution to the mean-field result, which leads 
to enhancement of the symmetrized shot noise in comparison to the mean-field result and a positive current CC for certain system 
parameters.

\begin{acknowledgments}

This work was supported by Projects of the National Basic Research Program of China (973 Program) under Grant No. 2011CB925603, 
and the National Science Foundation of China, Specialized Research Fund for the Doctoral Program of Higher Education (SRFDP) of 
China. The authors would like to thank the second referee for helpful comments in derivation of Eq.~(\ref{adp}). 

\end{acknowledgments}


\begin{thebibliography}{99}

\bibitem{Nazarov}{Yu.V. Nazarov, ed., {\em Quantum Noise in Mesoscopic systems}, NATO Science Series II, Vol. 97 (Kluwer, 
Dordrecht Boston London, 2003).}

\bibitem{Blanter}{Ya.M. Blanter and M. B\"uttiker, Phys. Rep. {\bf 336}, 1 (2000).}

\bibitem{Lu}{W. Lu, Z. Ji, L. Pfeiffer, K.W. West, and A.J. Rimberg, Nature (London) {\bf 423}, 422 (2003); J. Bylander, T. Duty 
and P. Delsing, Nature (London) {\bf 434}, 361 (2005).}

\bibitem{exp1}{B. Reulet, J. Senzier, and D.E. Prober, Phys. Rev. Lett. {\bf 91}, 196601 (2003); Yu. Bomze, G. Gershon, D. 
Shovkun, L.S. Levitov, and M. Reznikov, Phys. Rev. Lett. {\bf 95}, 176601 (2005); G. Gershon, Yu. Bomze, E.V. Sukhorukov, and M. 
Reznikov, Phys. Rev. Lett. {\bf 101}, 016803 (2008).}

\bibitem{Fujisawa}{T. Fujisawa, T. Hayashi, R. Tomita, Y. Hirayama, Science {\bf 312}, 551 (2006).}

\bibitem{exp2}{S. Gustavsson, R. Leturcq, B. Simovi\u{c}, R. Schleser, T. Ihn, P. Studerus, K. Ensslin, D.C. Driscoll, and A.C. 
Gossard, Phys. Rev. Lett. {\bf 96}, 076605 (2006); S. Gustavsson, R. Leturcq, T. Ihn, K. Ensslin, M. Reinwald, and W. 
Wegscheider, Phys. Rev. B {\bf 75}, 075314 (2007); E. V. Sukhorukov, A.N. Jordan, S. Gustavsson, R. Leturcq, T. Ihn, and K. 
Ensslin, Nature Phys. {\bf 3}, 243 (2007); S. Gustavsson, R. Leturcq, M. Studer, I. Shorubalko, T. Ihn, K. Ensslin, D.C. 
Driscoll, A.C. Gossard, Surface Science Reports {\bf 64}, 191 (2009).}

\bibitem{exp3}{A. Zazunov, M. Creux, E. Paladino, A. Cr\'epieux, and T. Martin, Phys. Rev. Lett. {\bf 99}, 066601 (2007).}

\bibitem{exp4}{C. Fricke, F. Hohls, W. Wegscheider, and R.J. Haug, Phys. Rev. B {\bf 76}, 155307 (2007); C. Flindt, C. Fricke, F. 
Hohls, T. Novotny, K. Netocny, T. Brandes, and Rolf J. Haug, Proc. Natl. Acad. Sci. USA {\bf 106}, 10116 (2009); C. Fricke, F. 
Hohls, N. Sethubalasubramanian, L. Fricke, and R.J. Haug, Appl. Phys. Lett. {\bf 96}, 202103 (2010).}

\bibitem{Levitov}{L.S. Levitov and G.B. Lesovik, JETP Lett. {\bf 58}, 230 (1993); D.A. Ivanov and L.S. Levitov, JETP Lett. {\bf 
58}, 461 (1993); L.S. Levitov, H.W. Lee, and G.B. Lesovik, J. Math. Phys. {\bf 37}, 4845 (1996).}

\bibitem{Shelankov}{A. Shelankov and J. Rammer, Europhys. Lett. {\bf 63}, 485 (2003).}

\bibitem{Levitov1}{L.S. Levitov and M. Reznikov, Phys. Rev. B {\bf 70}, 115305 (2004).}

\bibitem{Belzig}{W. Belzig and Y.V. Nazarov, Phys. Rev. Lett. {\bf 87}, 067006 (2001); {\bf 87}, 197006 (2001).}

\bibitem{Pilgram}{S. Pilgram, A. N. Jordan, E.V. Sukhorukov, and M. B\"uttiker, Phys. Rev. Lett. {\bf 90}, 206801 (2003); K. E. 
Nagaev, S. Pilgram, and M. B\"uttiker, Phys. Rev. Lett. {\bf 92}, 176804 (2004); S. Pilgram, K.E. Nagaev, and M. B\"uttiker, 
Phys. Rev. B {\bf 70}, 045304 (2004); M. Novaes, Phys. Rev. B {\bf 75}, 073304 (2007); Phys. Rev. B {\bf 78}, 035337 (2008).}

\bibitem{Taddei}{F. Taddei and R. Fazio, Phys. Rev. B {\bf 65}, 075317 (2002); H.-S. Sim and E.V. Sukhorukov, Phys. Rev. Lett. 
{\bf 96}, 020407 (2006); V. Giovannetti, D. Frustaglia, F. Taddei, and R. Fazio, Phys. Rev. B {\bf 74}, 115315 (2006); V. 
Giovannetti, D. Frustaglia, F. Taddei, and R. Fazio, Phys. Rev. B {\bf 75}, 241305 (2007).}

\bibitem{Lorenzo}{A. Di Lorenzo and Y.V. Nazarov, Phys. Rev. Lett. {\bf 93}, 046601 (2004); M. Kindermann, Phys. Rev. B {\bf 71}, 
165332 (2005).}

\bibitem{Bagrets}{D.A. Bagrets and Y.V. Nazarov, Phys. Rev. B {\bf 67}, 085316 (2003).}

\bibitem{Belzig1}{W. Belzig, Phys. Rev. B {\bf 71}, 161301 (2005).}

\bibitem{Kielich}{G. Kie{\ss}lich, P. Samuelsson, A. Wacker, and E. Sch\"oll, Phys. Rev. B 73, 033312 (2006); C.W. Groth, B. 
Michaelis, and C.W.J. Beenakker, Phys. Rev. B {\bf 74}, 125315 (2006); S.K. Wang, H. Jiao, F. Li, X.Q. Li, and Y.J. Yan, Phys. 
Rev. B {\bf 76}, 125416 (2007).}

\bibitem{Urban}{S. Welack, M. Esposito, U. Harbola, and S. Mukamel, Phys. Rev. B {\bf 77}, 195315 (2008); D. Urban and J. 
K\"onig, Phys. Rev. B {\bf 79}, 165319 (2009).}

\bibitem{Imura}{K.I. Imura, Y. Utsumi, and T. Martin, Phys. Rev. B {\bf 75}, 205341 (2007); Hai-Bin Xue, Y.-H. Nie, Z.-J. Li, 
J.-Q. Liang, J. Appl. Phys. {\bf 108}, 033707 (2010).}

\bibitem{Flindt}{C. Flindt, T. Novotn\'y, and A.-P. Jauho, Europhys. Lett. {\bf 69}, 475 (2005)}

\bibitem{Braggio}{A. Braggio, J. K\"onig, and R. Fazio, Phys. Rev. Lett. {\bf 96}, 026805 (2006); C. Flindt, T. Novotn\'y, A. 
Braggio, M. Sassetti, and A.-P. Jauho, Phys. Rev. Lett. {\bf 100}, 150601 (2008); C. Flindt, T. Novotn\'y, A. Braggio, and A.-P. 
Jauho, Phys. Rev. B {\bf 82}, 155407 (2010).}

\bibitem{Emary}{C. Emary, D. Marcos, R. Aguado, and T. Brandes, Phys. Rev. B 76, 161404 (2007); D. Marcos, C. Emary, T. Brandes, 
and R. Aguado, Phys. Rev. B {\bf 83}, 125426 (2011).}

\bibitem{Dong1}{Bing Dong, H.Y. Fan, X.L. Lei, N.J.M. Horing, J. Appl. Phys. {\bf 105}, 113702 (2009).}

\bibitem{Goldhaber}{D. Goldhaber-Gordon, H. Shtrikman, D. Mahalu, D. Abusch-Magder, U. Meirav and M.A. Kastner, Nature (London) 
{\bf 391}, 156 (1998); S.M. Cronenwett, T.H. Oosterkamp, and L.P. Kouwenhoven, Science {\bf 281}, 540 (1998).}

\bibitem{Franceschi}{S.De Franceschi, R. Hanson, W.G.van der Wiel, J.M. Elzerman, J.J. Wijpkema, T. Fujisawa, S. Tarucha, and 
L.P. Kouwenhoven, Phys. Rev. Lett. {\bf 89}, 156801 (2002); R. Leturcq, L. Schmid, K. Ensslin, Y. Meir, D.C. Driscoll, and A.C. 
Gossard, Phys. Rev. Lett. {\bf 95}, 126603 (2005)}

\bibitem{Paaske}{J. Paaske, A. Rosch, P. W\"olfle, N. Mason, C.M. Marcus and J. Nyg\aa rd, Nature Physics {\bf 2}, 460 (2006); M. 
Grobis, I.G. Rau, R.M. Potok, H. Shtrikman, and D. Goldhaber-Gordon, Phys. Rev. Lett. {\bf 100}, 246601 (2008).}

\bibitem{Zarchin}{O. Zarchin, M. Zaffalon, M. Heiblum, D. Mahalu, and V. Umansky, Phys. Rev. B {\bf 77}, 241303(R) (2008); T. 
Delattre, C. Feuillet-Palma, L.G. Herrmann, P. Morfin, J.-M. Berroir, G. F\`eve, B. Pla\c cais, D.C. Glattli, M.-S. Choi, C. Mora 
and T. Kontos, Nature Physics {\bf 5}, 208 (2009); Y. Yamauchi, K. Sekiguchi, K. Chida, T. Arakawa, S. Nakamura, K. Kobayashi, T. 
Ono, T. Fujii, and R. Sakano, Phys. Rev. Lett. 106, 176601 (2011)}

\bibitem{Hershfield}{S. Hershfield, Phys. Rev. B {\bf 46}, 7061 (1992).}

\bibitem{Yamaguchi}{F. Yamaguchi and K. Kawamura, J. Phys. Soc. Jpn. {\bf 63}, 1258 (1994).}

\bibitem{Ding}{G.-H. Ding and T.-K. Ng, Phys. Rev. B {\bf 56}, R15521 (1997).}

\bibitem{Dong2}{B. Dong and X.L. Lei, J. Phys.: Condens. Matter {\bf 14}, 4963 (2002).}

\bibitem{Lopez1}{R. L\'opez and D. S\'anchez, Phys. Rev. Lett. {\bf 90}, 116602 (2003)}

\bibitem{Sanchez}{D. S\'anchez and R. L\'opez, Phys. Rev. B {\bf 71}, 035315 (2005).}

\bibitem{Lopez2}{R. L\'opez, R. Aguado, and G. Platero, Phys. Rev. B {\bf 69}, 235305 (2004).}

\bibitem{Kubo}{T. Kubo, Y. Tokura, and S. Tarucha, Phys. Rev. B {\bf 83}, 115310 (2011).}

\bibitem{Meir}{Y. Meir and A. Golub, Phys. Rev. Lett. {\bf 88}, 116802 (2002).}

\bibitem{Golub}{A. Golub, Phys. Rev. B {\bf 72}, 075331 (2005).}

\bibitem{Bednorz}{A. Bednorz, and W. Belzig, Phys. Rev. Lett. {\bf 101}, 206803 (2008).}

\bibitem{Gogolin}{A.O. Gogolin and A. Komnik, Phys. Rev. B {\bf 73}, 195301 (2006);}

\bibitem{Meir2}{Y. Meir and N.S. Wingreen, Phys. Rev. Lett. {\bf 68}, 2512 (1992).}

\bibitem{Komnik}{A. Komnik and A.O. Gogolin, Phys. Rev. Lett. {\bf 94}, 216601 (2005)}

\bibitem{Gogolin2}{A.O. Gogolin and A. Komnik, Phys. Rev. Lett. {\bf 97}, 016602 (2006).}

\bibitem{Schmidt}{T.L. Schmidt, A.O. Gogolin, and A. Komnik, Phys. Rev. B {\bf 75}, 235105 (2007); T.L. Schmidt, A. Komnik, and 
A.O. Gogolin, Phys. Rev. B {\bf 76}, 241307(R) (2007).}

\bibitem{Schmidt2}{T.L. Schmidt, A. Komnik, and A.O. Gogolin, Phys. Rev. Lett. {\bf 98}, 056603 (2007).}

\bibitem{Schmidt3}{T.L. Schmidt and A. Komnik, Phys. Rev. B {\bf 80}, 041307(R) (2009); R. Avriller and A. Levy Yeyati, Phys. 
Rev. B {\bf 80}, 041309(R) (2009); S. Maier, T.L. Schmidt, and A. Komnik, Phys. Rev. B {\bf 83}, 085401 (2011).}

\bibitem{Coleman}{P. Coleman, Phys. Rev. B {\bf 29}, 3035 (1984).}

\bibitem{Langreth}{D.C. Langreth, in {\it Linear and Nonlinear Electron Transport in Solids, Nato ASI, Series B} vol. 17, Ed. J. 
T. Devreese and V. E. Van Doren (Plenum, New York, 1976).}

\bibitem{HBT}{M. Henny, S. Oberholzer, C. Strunk, T. Heinzel, K. Ensslin, M. Holland, and C. Sch\"onenberger, Science {\bf 284}, 
296 (1999); W.D. Oliver, J. Kim, R.C. Liu, and Y. Yamamoto, Science {\bf 284}, 299 (1999); H. Kiesel, A. Renz, and F. Hasselbach, 
Nature {\bf 418} 392, (2002); S. Oberholzer, E. Bieri, C. Sch\"onenberger, M. Giovannini, and J. Faist, Phys. Rev. Lett. {\bf 
96}, 046804 (2006).}

\bibitem{Anantram}{M.P. Anantram and S. Datta, Phys. Rev. B {\bf 53}, 16 390 (1996); T. Martin, Phys. Lett. A {\bf 220}, 137 
(1996); J. Torres and T. Martin, Eur. Phys. J. B {\bf 12}, 319 (1999); F. Taddei and R. Fazio, Phys. Rev. B {\bf 65}, 134522 
(2002).}

\bibitem{Cottet} {A. Cottet, W. Belzig, and C. Bruder, Phys. Rev. Lett. \textbf{92}, 206801 (2004); A. Cottet, W. Belzig, and C. 
Bruder, Phys. Rev. B \textbf{70}, 115315 (2004).}

\bibitem{Martin}{A.M. Martin and M. B\"uttiker, Phys. Rev. Lett. {\bf 84}, 3386 (2000).}

\bibitem{Dong4}{B. Dong, X.L. Lei, and N.J.M. Horing, Phys. Rev. B {\bf 80}, 153305 (2009).}

\bibitem{McClure}{D.T. McClure, L. DiCarlo, Y. Zhang, H.-A. Engel, C.M. Marcus, M.P. Hanson and A.C. Gossard, Phys. Rev. Lett. 
{\bf 98}, 056801 (2007).}

\end{thebibliography}
\end{document}